# S-Shaped Flow Curves of Shear Thickening Suspensions: Direct Observation of Frictional Rheology

Zhongcheng Pan[1], Henri de Cagny[1], Bart Weber[1], Daniel Bonn[1]

[1]*Van der Waals-Zeeman Institute, IoP, Science Park 904, Amsterdam, Netherlands*

We study the rheological behavior of concentrated granular suspensions of simple spherical particles. Under controlled stress, the system exhibits an S-shaped flow curve (stress vs. shear rate) with a negative slope in between the low-viscosity Newtonian regime and the shear thickened regime. Under controlled shear rate, a discontinuous transition between the two states is observed. Stress visualization experiments with a novel fluorescent probe suggest that friction is at the origin of shear thickening. Stress visualization shows that the stress in the system remains homogeneous (no shear banding) if a stress is imposed that is intermediate between the high and low-stress branches. The S-shaped shear thickening is then due to the discontinuous formation of a frictional force network between particles upon increasing the stress.

The phenomenon of shear thickening is important for many industrial applications [1] and exists in a wide range of systems, including wormlike micelle solutions [2-4], cornstarch [5-8] and colloidal [9,10] and non-colloidal suspensions [1,7,11-12]. Granular suspensions made of spherical particles dispersed in a Newtonian liquid are arguably the simplest of these systems; nonetheless their rheological behavior is very rich. If the particles and solvent are not perfectly density matched, such suspensions will exhibit a yield stress and pronounced shear thinning [12]; in addition the measured viscosity can be significantly affected by particle migration [1]. For a perfectly density matched system without migration, besides a Newtonian flow regime, both continuous shear thickening and discontinuous shear thickening can be observed depending on the volume fraction of particles [13,14]. The resulting difficulty in predicting the flow behavior of a given suspension hampers our understanding the rheological behavior of granular suspensions. This is unfortunate since the handling and transport of granular materials in general is responsible for a significant fraction of the world energy consumption [15].

From a fundamental point of view, shear thickening is of great interest since it is a remarkable exception to the general rule that most complex fluids organize themselves in flow to minimize the flow resistance. Shear thickening is the opposite and often described as a shear-induced jamming transition [1,7,8]; however, other mechanisms are also under debate [6,16]. Consequently, to precisely predict the thickening behavior remains a challenge. For instance, the assertion that shear thickening is due to the inertia of the particles implies that the thickening happens at a Stokes number St $\approx$ 1 as observed in some simulations [17-18], in stark contrast to St $\approx 10^{-3}$ observed in experiments [1,19]. Recent simulations, on the other hand, suggest that inertia is not important for shear thickening [13,20]. It is even harder to estimate which systems will shear thicken and whether the thickening is monotonic or not [1]. Theoretical approaches to quantitatively describe the thickening behavior utilize the perspective of either hydrodynamic interactions or geometric and steric constraints [16]; however, neither approach is completely satisfying. A recent simulation that considers frictional contact between hard spherical particles seems to unveil the underlying physics behind shear thickening [14,21]. When particle-particle frictional forces are incorporated into the hydrodynamic description, the transition from continuous to discontinuous thickening and even a non-monotonic thickening at high volume fractions can be qualitatively predicted [13,20]. The significance of friction is evident from models that take a finite-range interaction into account [13], suggesting that shear thickening arises due to frictional contacts between particles when the finite-range particle-particle repulsion is overcome by the applied shear stress. Despite an emerging consensus that this is the case, few experimental observations are available [16,22] and direct evidence is urgently needed.

In this Letter, we present such evidence. We first show by that concentrated suspensions surprisingly exhibit an S-shaped flow curve under controlled shear stress with a hysteresis that depends on the rate at which stress sweeps are performed. This S-shaped flow curve is only observed at the volume fractions where by controlling the shear rate discontinuous shear thickening occurs. The stable flowing thickened states on the other hand indicate that jamming is not a prerequisite for observing discontinuous shear thickening. For shear-thickening micellar systems, S-shaped flow curves are associated with shear banding [2-4], and the question arises whether the suspensions show analogous behavior. We therefore investigate the local stresses during thickening using a novel fluorescent probe whose fluorescence intensity depends on the imposed stress between particles. These measurements reveal that our system remains homogeneous and suggest that the origin of the shear thickening is dynamical stress-induced frictional contact proliferation between the particles of our system.

The granular suspensions used in our experiments are made by dispersing neutrally buoyant non-Brownian particles in a Newtonian solvent, water. We use PMMA particles with diameter $d = 10$ μm and density $\rho = 1.19$ g/cm$^3$. To avoid sedimentation or creaming, we prepare density matched suspensions by adding Sodium Iodide (NaI, Sigma Aldrich) to the water in order to adjust the density to that of the particles. Due to density matching, we can set the volume fractions by calculating $\varphi = m_p/m$ with $m_p$ and $m$ the mass of granular particles and the total mass of the suspension respectively. The initial volume fraction $\varphi$ is varied from 55%-59%. For density-matched suspensions, no contacts induced by gravity exist and normal forces are only caused by shear [1].

The rheological measurements are performed by a rheometer (Anton Paar MCR300) with a small-gap Couette geometry: a rotating inner cylinder of 27 mm in diameter and a fixed outer cup diameter of 29 mm, leading to a gap of 1 mm. This gap size is around 60-100 times the particle diameter so that finite-size effects are negligible. We verified that the stress distribution in the gap is uniform compared to that in the wide-gap Couette geometry that is often used for these suspensions [1,12]; no strong particle migration effects occur when the suspension is measured over long time: variations in viscosity (volume fractions) are smaller than a few percent. For controlling both shear stress and shear rates, the sweep rates are set at 10 s/data point, 30 points/decade unless specified otherwise.

To visualize the flow behavior of the suspensions, we use a fast Confocal Microscope (Zeiss Pascal Live) coupled with a DSR 301 rheometer head. We use a cone-plate geometry CP50-1 (50 mm/1°) with a gap of 0.102 mm, but replaced the usual bottom plate by a transparent glass microscope slide. Because the confocal is an inverted microscope, the sample in the cone-plate geometry can be directly visualized by making microscopy images through the glass slide while the sample flows. We apply a novel technique to visualize the local stress, we use the fluorescent stress probe molecule 9-(2,2-Dicyanovinyl) julolidine (DCVJ) dissolved in the aqueous phase; DCVJ belongs to a class of rigidochromic molecular rotors based on Twisted Intramolecular Charge Transfer (TICT) states [23,24] and is sensitive to the normal stress between particles: the higher the stress, the higher the fluorescence intensity it emits [23]. To our knowledge, this is the first report that uses such stress probes in a non-Newtonian liquid to detect frictional contacts between particles.

The concentration-dependent shear thickening behavior measured in the Couette geometry is shown in **Fig**. 1. Upon increasing the stress, we first observe a Newtonian flow behavior at low stress in agreement with [12]. Here, some fluctuations may be due to slight particle migration effects or a slight density mismatch as a result of varying lab temperatures. Next, clear thickening behavior at higher stresses is observed: continuous shear thickening occurs for low volume fractions ( $\varphi \leq 56.5\%$ ), becoming more pronounced with increasing φ. Surprisingly, a fraction above a threshold value $\varphi \approx 56.5\%$ (a value very similar to that in [13,20]), the flow curves display continuous shear thickening first, followed by an S-shaped flow curve. When $\varphi$ exceeds 58%, the continuous thickening weakens and the viscous Newtonian regime and the high-viscosity thickened regime are only connected by an intermediate part with a negative slope. The thickened states in both S-shaped and discontinuous shear thickening are reversible, indicating that complete jamming does not occur here and thus is not necessary for the discontinuous shear thickening. In jamming, the viscosity would become infinite [6] and the system cannot flow without (particle) inhomogeneity and fracture [13].

**Fig**. 1 shows that the onset stress for the S-shaped curve decreases with increasing φ, which is different from the shear rate-controlled rheology reported in [1], where the onset stress varies only weakly with φ. The onset shear rate (stress) can be estimated by considering dilatancy [25] that causes a non-equilibrium osmotic pressure (particle pressure) $\Pi \approx \eta\dot{\gamma}/(1 - \varphi/\varphi_m)^2$ with $\varphi_m$ the maximum volume fraction [26,27]. A simple estimate can be made by equating this pressure to the Laplace pressure given as $\gamma/d$,

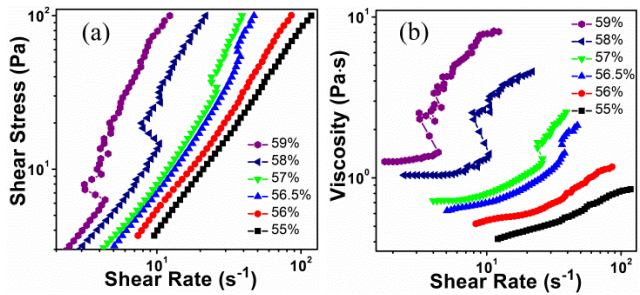

FIG.1. Flow curves under controlled shear stress: (a) shear stress vs. shear rate and (b) viscosity vs. shear rate for granular suspensions with volume fractions varying from 55% to 59%. The flow curves are taken when the suspensions are sheared over a long time.

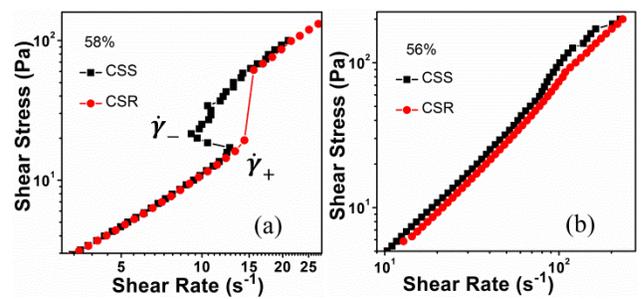

FIG.2. Comparison between flow curves (stress vs. shear rate) obtained from controlling shear stress (CSS, black squares) and controlling shear rate (CSR, red circles). The volume fractions are fixed at 58% (a) and 56% (b).

where $\gamma$ is the surface tension and $d$ is the particle diameter. With $\eta \approx 1\, Pa \cdot s$, $\varphi_m \approx 0.63$ and $\gamma \approx 0.02 N/m$, we obtain a critical shear rate $\dot{\gamma}_c \approx 12.6\, s^{-1}$ in good agreement with the onset shear rate for S-shaped flow curve at $\varphi = 58\%$ (**Fig.** 1).

We now investigate the S-shaped flow curve in more detail for concentrated suspensions with $\varphi$ fixed at 58%. **Fig.** 2a shows the difference between shear stress and shear rate controlled experiments. While under stress control we obtain the S-shaped flow curve as above, under shear rate control a discontinuous jump in stress is observed: by controlling shear stress, measurements can be performed beyond the onset of sudden shear thickening [28]. Both flow curves exhibit the same Newtonian regime and shear thicken at almost the same shear rate ($\dot{\gamma}_+$). Under shear rate control however, the stress abruptly jumps to a higher value corresponding to the thickened state, while in the stress controlled experiments the system has to pass through the S-shaped curve characterized by a second critical shear rate $\dot{\gamma}_-$ at which the thickened state is reached. The two types of flow curves are again identical in the thickened state. In addition, the volume fraction where S-shaped flow curve appears coincides with the appearance of discontinuous shear thickening in controlled shear rate experiments; also as shown in **Fig.** 2b, no quantitative difference is seen between shear stress controlled and shear rate controlled experiments at a low volume fraction $\varphi = 56\%$.

A hysteresis, similar to that observed in cornstarch suspensions [29], is observed when we impose up-and-down stress sweeps (10 s/data point) on the sample. The S-shaped flow curve is observed in both upward and downward shear stress sweeps (**Fig.** 3a). The hysteresis region in **Fig.** 3a can be roughly described as a rectangle with two vertices on the Newtonian branch (at $\dot{\gamma}_+ \approx 12.5\, s^{-1}$ and at $\dot{\gamma}_- \approx 9.5\, s^{-1}$) and two vertices on the shear thickened branch. The rate at which stress sweeps are performed determines the flow curves; **Fig.** 3b shows that at a rate of 40 s/data point, the negative slope sides of the rectangle approach each other and the hysteresis disappears, while the S-shape in the flow curve remains. **Fig.** 3a shows that up-and-down shear rate sweeps also result in hysteresis loops, again in accordance with shear stress controlled rheology in that the onset shear rate is identical to that for observing the S-shaped flow curve.

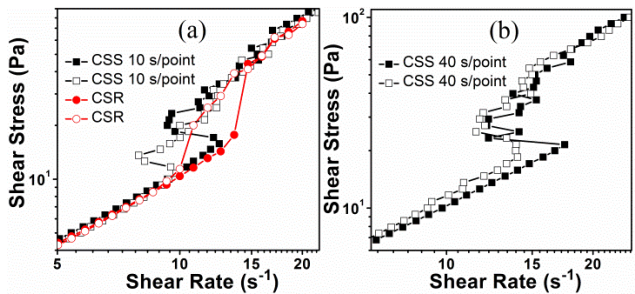

FIG.3. Up-and-down flow curves displaying hysteresis: stress vs. shear rates at stress sweep rates of (a) 10 s/point and (b) 40 s/point. Filled symbols are for increasing stress and open symbols are for decreasing stress sweeps. Black symbols are under controlled shear stress (CSS), red symbols under controlled shear rate (CSR).

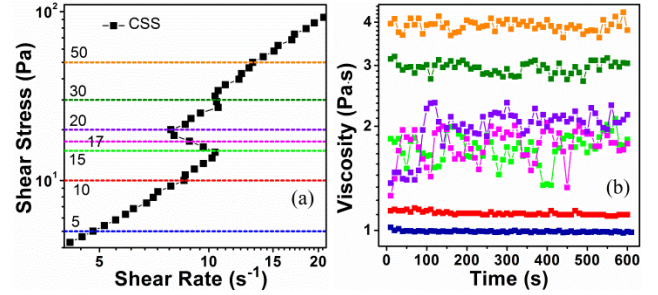

FIG.4. (a) Flow curve on which the different constant stress levels are indicated: 5, 10 Pa for the Newtonian state, 15, 17, 20 Pa for the intermediate state and 30, 50 Pa for the thickened state. The corresponding viscosities are shown as a function of time in (b). The symbols in (b) are in accordance with the dashed lines in (a) of the same color.

This kind of hysteresis is often attributed to stress heterogeneity and is believed to be analogous to the hysteresis accompanying coexistence of two phases in a first-order phase transition [29]: the negative slope of stress vs. shear rate cannot reflect the viscosity of a homogeneous system; in general it signals a linear instability of such a flow, which for systems like wormlike micelles results in shear banding.

To investigate what happens in this part of the S-shaped flow curve, in **Fig.** 4a we show the result of a series of constant shear stress experiments, taken at varying locations along the flow curve. **Fig.** 4b shows that when constant stress is imposed in either the Newtonian state or the thickened state, the viscosities stay almost constant over a period of 10 min suggesting two stable flowing states (no jamming). However, under constant controlled stress in the intermediate state, the viscosities in our sample fluctuate between 1.3 and 2.5 $Pa \cdot s$. Comparing with **Fig.** 4a, this corresponds roughly to the viscosity at $\dot{\gamma}_+$ and $\dot{\gamma}_-$ for which we find 1.5 and 2.7 $Pa \cdot s$ from the flow curve, respectively.

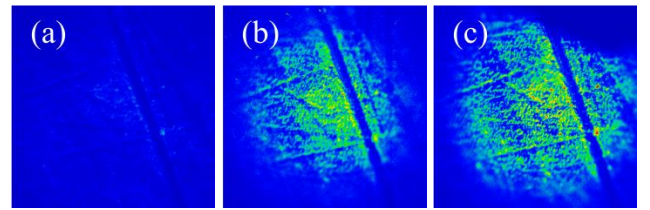

FIG.5. Representative fluorescent images with the focal plane positioned at the surface of a transparent glass slide on top of which one drop of DCVJ aqueous solution is loaded. A PMMA bead is pressed on the glass slide with the following normal stress levels: (a) 0 mN, (b) 131 mN, (c) 240 mN.

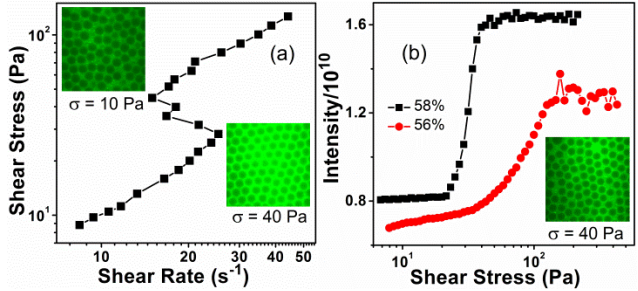

FIG.6 (a) Flow curve: stress vs. shear rate for the suspension at $\varphi = 58\%$ in the confocal cone-plate rheometer. Inset: two confocal images taken at a magnification of 63x are typical for low and intermediate stress respectively. The images taken at other magnifications show similar results. (b) The fluorescence intensity changes over the controlled stress sweep (30 points per decade and 3 s per data point). Inset: local fluorescence response of DCVJ.

These observations could suggest that the Newtonian and the shear-thickened state coexist here. Theory [30] indicates that in this case an S-shaped flow curve should be associated with shear banding in the vorticity direction under controlled stress: the system separates into bands of different stresses (thus different viscosities) that pile up along the vorticity direction.

To directly investigate the vorticity banding hypothesis we need to be able to distinguish parts of the system that have different shear stresses but the same shear rate, which is far from obvious. **Fig.** 5 however shows that the fluorescence emission of DCVJ, present in the interstitial fluid between the particles can be turned on by increasing the normal stress between the particles. The onset of shear thickening has often been associated with the emergence of normal stresses between particles, and consequently the DCVJ can be used as a local stress sensor [23]. **Fig.** 6a shows that the S-shaped flow curve is reproduced in the cone-plate geometry, albeit with a slightly higher onset shear rate (stress), which is perhaps due to the change in geometry leading to a different dilatancy effect. By focusing the fast confocal microscope in a layer 20 $\mu m$ above the glass slide we find,, that the stress field in the system remains uniform even for shear stresses corresponding to the negative slope part of the S-shaped curve (**Fig.** 6a). This happens in spite of the fact that the overall fluorescence intensity does increase with increasing the imposed shear stress (**Fig.** 6b), showing for the first time that these molecules can be used to monitor flows for non-Newtonian fluids.

**Fig.** 6b records the total fluorescence intensity over the stress sweep process for the concentrated suspension at $\varphi = 58\%$ as well as for the diluted suspension at $\varphi = 56\%$. In both cases, the fluorescence first increases mildly in the Newtonian regime upon increasing stress, indicating no effective frictional contact network between particles. In the thickening regime the fluorescence for $\varphi = 56\%$ suspension increases faster but still continuously. This is very much consistent with the continuous shear thickening behavior of **Fig.** 1, suggesting friction is indeed at the origin of shear thickening. For the $\varphi = 58\%$ suspension, the fluorescence increases abruptly and almost discontinuously; the S-shaped curve is then caused by a sudden mobilization of a frictional force network between particles: due to the controlled stress, the only way that the system can respond to the increased effective viscosity is to decrease the shear rate, causing the negative slope part of the S-shaped curve. For the highest stresses, the fluorescence in both cases saturates.

Recent simulation results [13,20] also reveal the existence of an S-shaped flow curve and similarly to what is observed here, the authors argue that the underlying cause is a frictional network that is a monotonic function of the intensity of stress chains. The interpolation between the two states (Newtonian state and thickened state) is then simply a stress-based mixing rule. In this case, a macroscopic "phase separation" (i.e. vorticity banding) does not show up, in line with our observations. The viscosity (stress) fluctuation shown in **Fig.** 4b is likely due to building and release of local stress in the formation of the percolating frictional network.

In summary, we experimentally study the S-shaped flow curve of concentrated granular suspensions and for the first time directly observe frictional rheology, in agreement with recent theory and (stress controlled) simulations. We conclude that friction is at the origin of shear thickening. For the S-shaped flow curve, when a constant stress is applied in between the high and low-stress branches, our visualization experiments suggest that the flow remains homogeneous during the transition from the Newtonian to the shear thickened state, in line with the idea that a frictional force network forms dynamically in the flow; once this network percolates, the shear-thickened regime is reached.